\documentclass[seceq,supplement]{ptptex}

\usepackage{graphicx} 
\title{%        %You can use \\ for explicit line-break.
On the Origin of the Trigger-Angle Dependence of the Ridge  Structure%
}

\author{%       %Use \scshape for the family name.
Yogiro \textsc{Hama}$^{1,}$\footnote{Speaker}, 
Rone P.G. \textsc{Andrade}$^1$, 
Fr\'ed\'erique \textsc{Grassi}$^1$ 
and Wei-Liang \textsc{Qian}$^2$%
}

\inst{%     %Affiliation, neglected when [addenda] or [errata].
$^1$Instituto de F\'{\i}sica, Universidade de S\~ao Paulo, 
 S\~ao Paulo, Brazil\\ 
$^2$Departamento de F\'{\i}sica, Universidade Federal de 
  Ouro Preto, Ouro Preto-MG, Brazil
}

\abst{%         %This abstract is neglected when [addenda] or [errata].
The origin of the trigger-angle dependence of the ridge 
structure in two-hadron long-range correlations, as observed 
at RHIC, is discussed as due to an interplay between the 
elliptic flow caused by the initial state global geometry and flow produced by fluctuations. 
}

\begin{document}

\maketitle

\section{Introduction} 
One of the most striking results in relativistic heavy-ion collisions at RHIC and LHC, is the existence of  structures  in the two-particle correlations  \cite{R2,star3,R1,Ralice,Rcms,Ratlas} plotted as function of the pseudorapidity difference $\Delta\eta$ and the angular 
spacing $\Delta\phi$. The so-called ridge has a narrow  $\Delta \phi$ located around zero and a long $\Delta \eta$  extent. The other structure located opposite has a single or 
double hump in $\Delta \phi$. 

In a previous work,\cite{jun} we got the ridge structure in 
a purely hydrodynamic model. What is essential to producing 
ridges in hydrodynamic approach are: i) Event-by-event fluctuating initial conditions (IC); and besides, ii) Very bumpy tubular structure in the IC. Our code NeXSPheRIO uses this kind of IC, produced by NEXUS event generator,\cite{nexus} connected to SPheRIO hydro code.\cite{spherio} In our previous studies on ridge, by using 3D NeXSPheRIO code, we obtained some of the  experimentally known properties such as 
\begin{itemize} 
 \item centrality 
  dependence,\cite{ismd09,sqm,bnl}  
 \item trigger-direction dependence in non-central 
  windows,\cite{ismd09,sqm,bnl,ismd10r}  
 \item $p_T$ dependence,\cite{sqm,bnl}  
 \end{itemize}  
 
However, {\it what is the origin of ridges?} In order to understand the dynamics of ridge formation, we studied  carefully what happens in the neighborhood of a peripheral  high-energy tube, introducing what we call 
{\it boost-invariant one-tube model.}\cite{ismd09}   

\section{In-plane/out-of-plane effect} 

\begin{figure}
 \centerline{\includegraphics[width=12.cm]{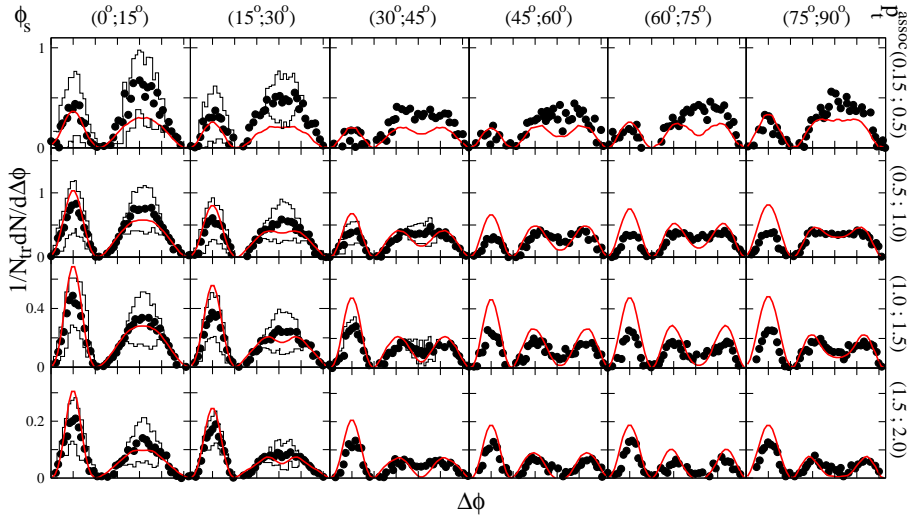}}
  \caption{Comparison of NeXSPheRIO results (solid curves) 
   on the trigger-angle dependence of two-particle 
   correlation, as function of $\Delta\phi$, with 
   data points~\cite{in_out_1}.} 
  \label{fig:1}
\end{figure}

In non-central collisions (20-60\% centrality), data have 
been obtained, fixing the azimuthal angle of the trigger 
($\phi_S$) with respect to the event plane.\cite{in_out_1,in_out_2} 
As shown in Fig.~\ref{fig:1}, the ridge structure in $\Delta\phi$ 
depends on the trigger direction, especially on the away-side. It changes  continuously from one-broad-peak structure at $\phi_S=0$ (in-plane trigger), to double-peak structure at $\phi_S=\pi/2$ (out-of-plane trigger). 
The change is particularly manifest 
for larger transverse momentum of the associated particle. We tried 
to see whether this behavior can be obtained with our NeXSPhe\-\hfilneg\ 

\noindent RIO code. 
The results are shown in Fig.~\ref{fig:1}, with solid lines. Here,  
we emphasize that these are results of no-parameter computations, having been all the parameters fitted previously using the single-particle 
$\eta$ and $p_T$ distributions. So, we think NeXSPheRIO code can reproduce the data quite reasonably. 

%\begin{wrapfigure}{r}{6.6cm} 
% %\centerline{\includegraphics[width=6.cm]{in-out_final.eps}}
% \caption{Two-particle correlation as function of %$\Delta\phi$ 
% for in-plane (one away-side peak) and out-of-plane %trigger (double away-side peak).}
% \label{fig:2}
%\end{wrapfigure}
Now, {\it how this effect is produced?} Trying to clarify the 
origin of the effect, we again used the boost-invariant one-tube model, now adapted to non-central collisions. In Ref.~\citen{rone}, we took as 
the background the average energy-density distribution obtained with  NeXSPheRIO, which now has an elliptical shape. A randomly distributed peripheral tube is put on top of this as our initial conditions. As shown there 
%and reproduced in Fig.~\ref{fig:2}, 
the two-particle correlation obtained with this model does reproduce the in-plane/out-of-plane effect, as shown by data.\footnote{See more details in Ref.~\citen{rone}.}
 
However, the mechanism of this effect is still not transparent. In order to make it clearer, we try the following analytical model, which is valid if the amplitude of fluctuations is small enough:  
\begin{equation} 
  \frac{dN}{d\phi}(\phi,\phi_t)
  =\frac{dN_{bgd}}{d\phi}(\phi) 
  +\frac{dN_{tube}}{d\phi}(\phi,\phi_t), 
 \label{eq:1}  
\end{equation} 
where 
\begin{eqnarray} 
 \frac{dN_{bgd}}{d\phi}(\phi)&=&\frac{N_b}{2\pi}
    (1+2v_2^b\cos(2\phi))\quad\quad \hbox{and}
    \label{eq:2}\\  
 \frac{dN_{tube}}{d\phi}(\phi,\phi_t)&=&\frac{N_t}{2\pi} 
    \sum_{n=2,3}2v_n^t\cos(n[\phi-\phi_t])\,.
    \label{eq:3}     
\end{eqnarray} 
The azimuthal angle $\phi$ is measured with respect to the event plane 
and $\phi_t$ is the location of the randomly distributed tube. 

The two-particle correlation is given by 
\begin{equation} 
 \langle\frac{dN_{pair}}{d\Delta\phi}(\phi_s)\rangle 
  =\langle\frac{dN_{pair}}
   {d\Delta\phi}(\phi_s)\rangle^{proper} 
  -\langle\frac{dN_{pair}}
   {d\Delta\phi}(\phi_s)\rangle^{mixed},  
\end{equation} 
where, in one-tube model, 
\begin{eqnarray} 
 \langle\frac{dN_{pair}}{d\Delta\phi}\rangle^{proper}\!
  &=&\int\frac{d\phi_t}{2\pi}f(\phi_t)
     \frac{dN}{d\phi}(\phi_s,\phi_t)
     \frac{dN}{d\phi}(\phi_s+\Delta\phi,\phi_t) 
     \quad\quad\hbox{and}\\  
 \langle\frac{dN_{pair}}{d\Delta\phi}\rangle^{mixed}\!
  &=&\!\!\int\frac{d\phi_t}{2\pi}f(\phi_t)
         \int\frac{d\phi_t'}{2\pi}f(\phi_t') 
     \frac{dN}{d\phi}(\phi_s,\phi_t)
     \frac{dN}{d\phi}(\phi_s\!+\!\Delta\phi,\phi_t'). 
\end{eqnarray} 
Here, $\phi_s$ is the trigger angle ($\phi_s=0$ for in-plane and 
$\phi_s=\frac{\pi}{2}$ for out-of-plane trigger) and $f(\phi_t)$ is 
the distribution function of the tube. We will take $f(\phi_t)=1$, 
for simplicity. Notice that in the mixed events, integrations both 
over $\phi_t$ and $\phi_t'$ are required, whereas only one 
integration, over $\phi_t$ is enough for proper events in the 
averaging procedure. This difference becomes essential. 

Using our simplified parametrization, (\ref{eq:1}), (\ref{eq:2}), (\ref{eq:3}) and, by averaging over events, the two-particle correlation for the in-plane trigger is  given as 
\begin{equation}
 \langle\frac{dN_{pair}}{d\Delta\phi}\rangle^{proper}_{in}\,\,
  =\frac{<N_b^2>}{(2\pi)^2}(1+2v_2^b)
                    (1+2v_2^b\cos(2\Delta\phi))
  +(\frac{N_t}{2\pi})^2
     \sum_{n=2,3}2(v_n^t)^2\cos(n\Delta\phi)
 \label{eq:7}
\end{equation}
and       
\begin{equation}       
 \langle\frac{dN_{pair}}{d\Delta\phi}\rangle^{mixed}_{in}\!
  =\frac{<N_b>^2}{(2\pi)^2}(1+2v_2^b)
                    (1+2v_2^b\cos(2\Delta\phi))\,.
 \label{eq:8}                    
\end{equation}
Observe the difference between the factors multiplying the background terms of the proper- and the mixed-event correlations. So, by subtracting (\ref{eq:8}) from  (\ref{eq:7}), the resultant in-plane correlation is 
\begin{eqnarray} 
 \langle\frac{dN_{pair}}{d\Delta\phi}\rangle_{in-plane} 
  &=&\frac{<N_b^2>-<N_b>^2}{(2\pi)^2}
     (1+2v_2^b)(1+2v_2^b\cos(2\Delta\phi))\nonumber\\ 
  &+&(\frac{N_t}{2\pi})^2 
      \sum_{n=2,3}2(v_n^t)^2\cos(n\Delta\phi)\,, 
 \label{eq:9} 
\end{eqnarray} 
{\it i.e.}, if the multiplicity fluctuates {\it the 
background elliptic flow does contribute to the correlation.}  

\begin{figure} 
  {%\vspace*{-.cm} 
  \begin{minipage}[h]{0.33\textwidth}
    \vspace*{-1.5cm} 
    \centerline{%\vspace*{-3.cm}
    \includegraphics[width=4.2cm]{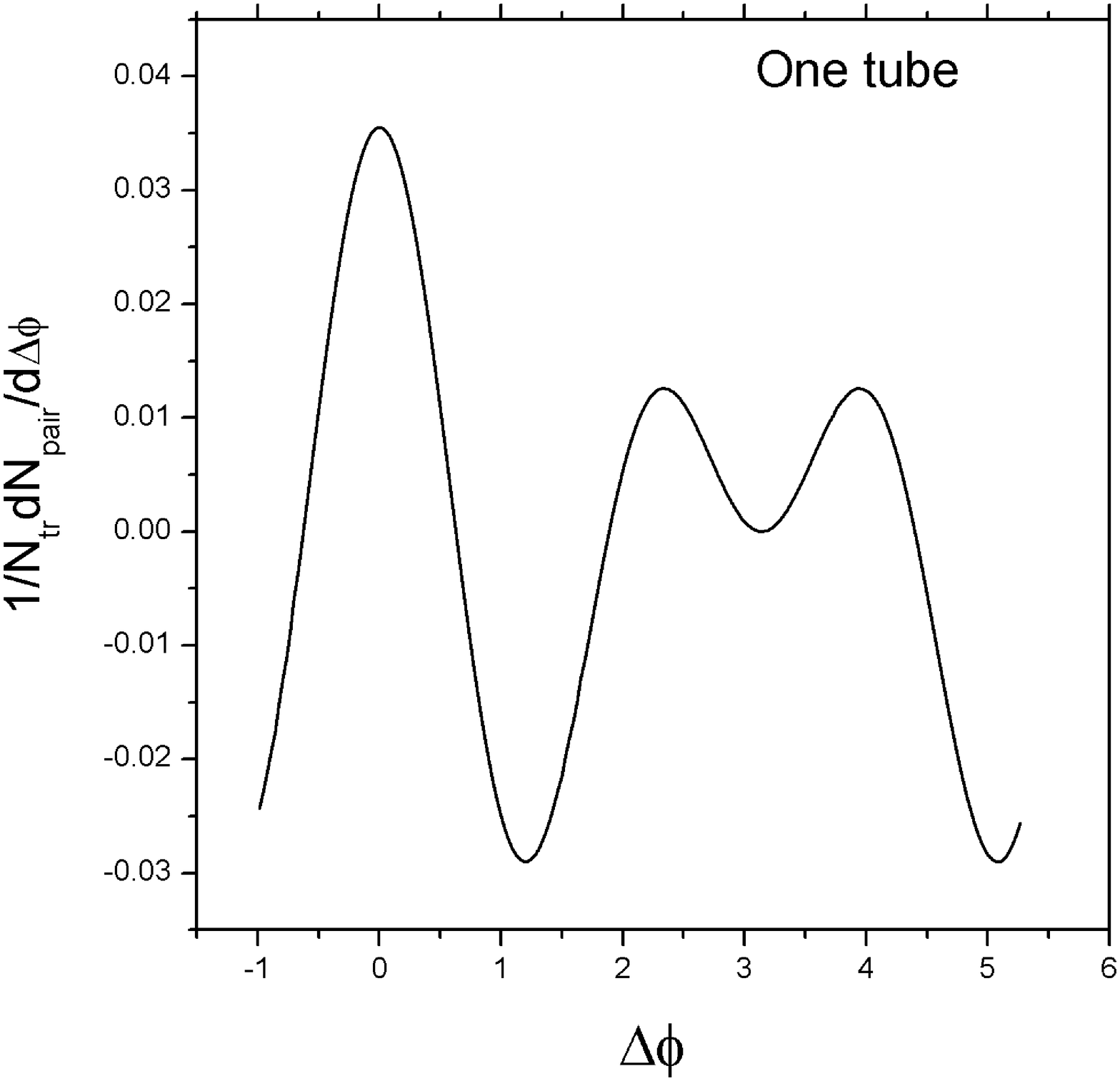}}
  \end{minipage}  } 
  \begin{minipage}[h]{0.335\textwidth}
    \centerline{\includegraphics[width=5.cm]{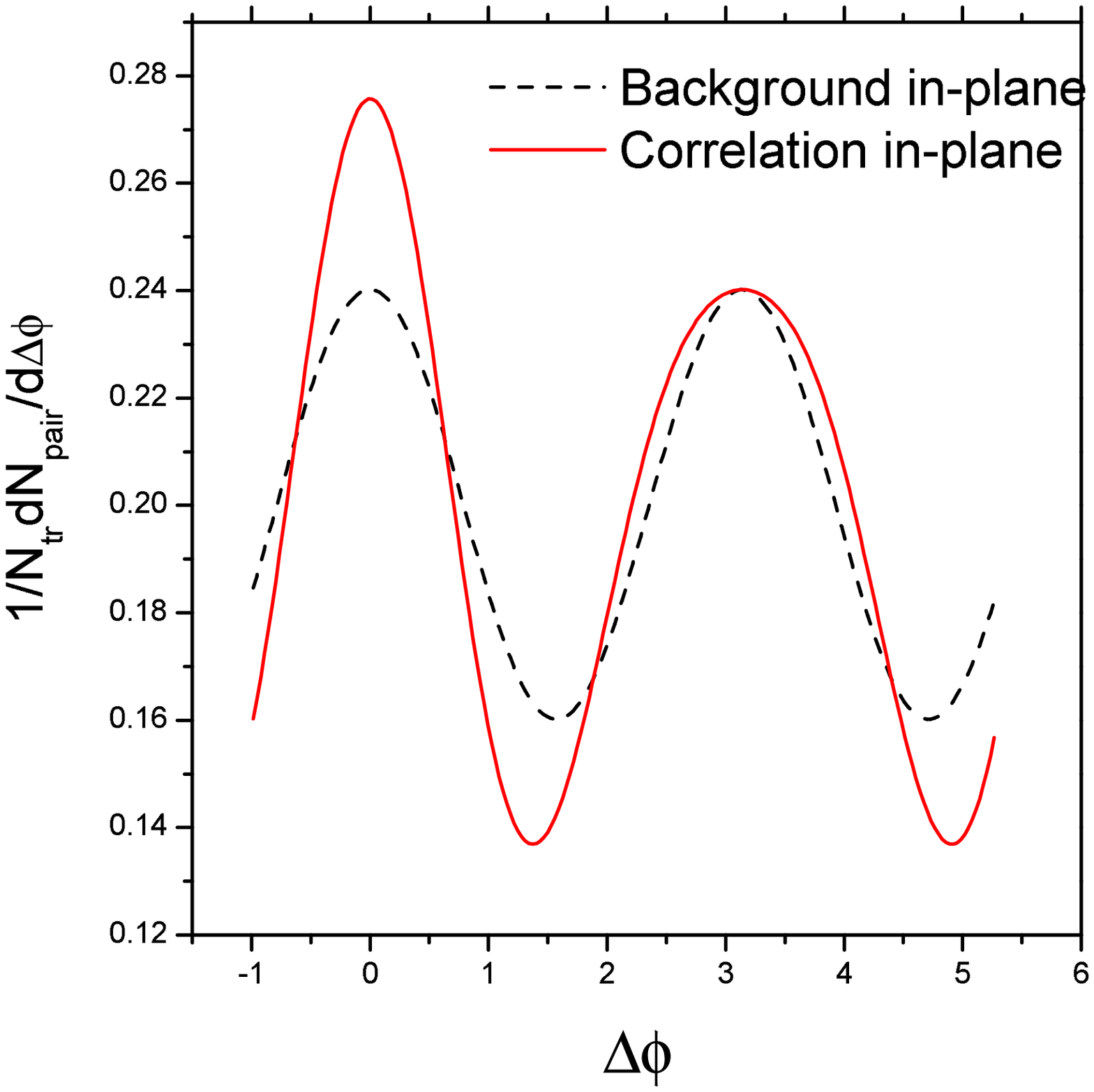}}
  \end{minipage}  
  \begin{minipage}[h]{0.335\textwidth}
    \centerline{\includegraphics[width=5.cm]{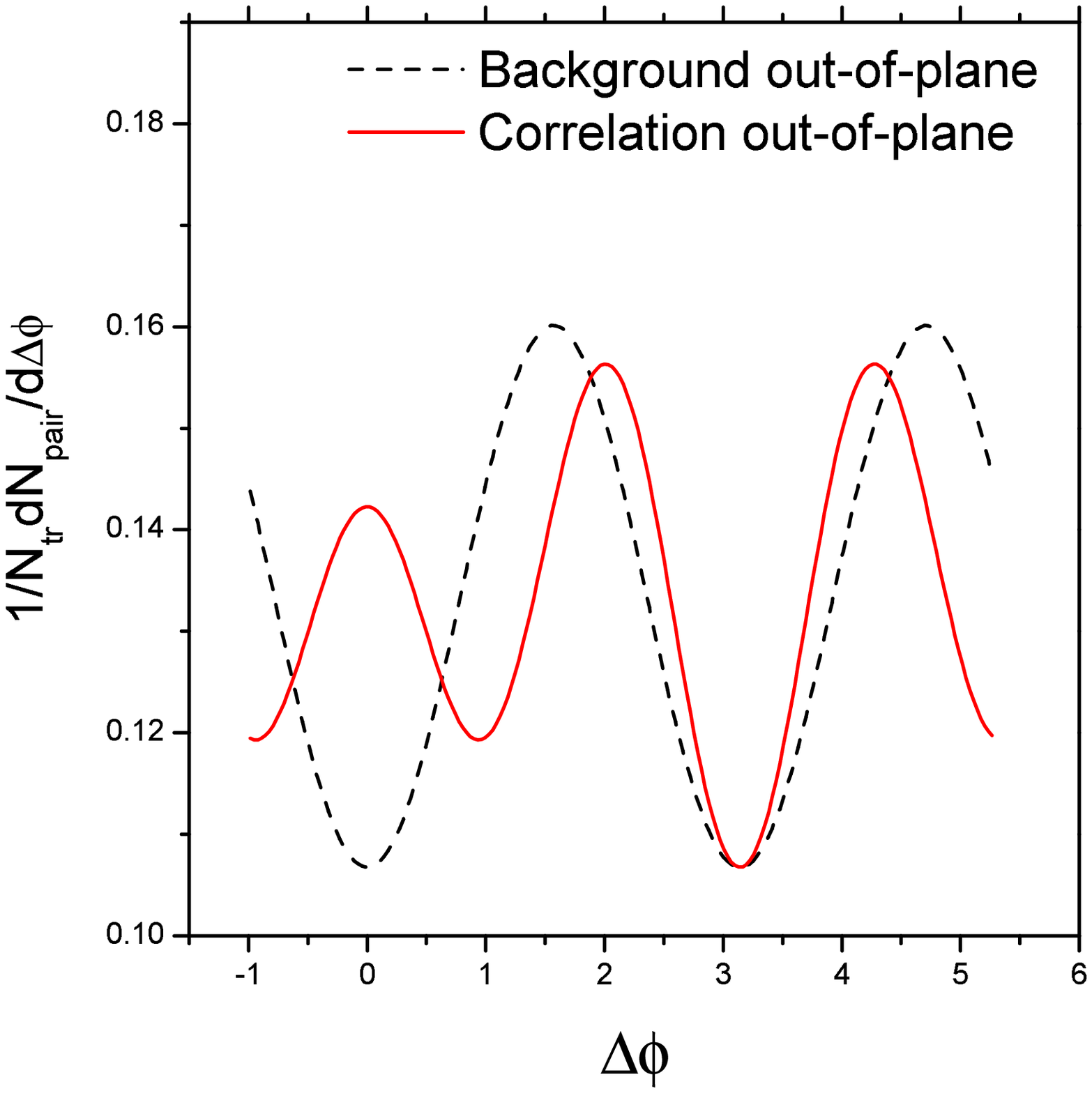}}
  \end{minipage}  
 \vspace*{-2.cm}            
 \caption{Plots of two-particle correlation. From the left to the 
  right, i) the one-tube contribution; ii) the one from the 
  background (dashed line) and the total one (solid line) for 
  in-plane triggers, as given by (\ref{eq:9}); and iii) the 
  corresponding ones for the out-of-plane triggers (\ref{eq:10}).} 
 \label{fig:2}           
\end{figure}

Similarly, the out-of-plane correlation is given as 
\begin{eqnarray} 
 \langle\frac{dN_{pair}}{d\Delta\phi}\rangle_{out-of-plane} 
  &=&\frac{<N_b^2>-<N_b>^2}{(2\pi)^2}
     (1-2v_2^b)
     (1-2v_2^b\cos(2\Delta\phi))\nonumber\\ 
  &+&(\frac{N_t}{2\pi})^2 
      \sum_{n=2,3}2(v_n^t)^2\cos(n\Delta\phi)\,.   
 \label{eq:10} 
\end{eqnarray} 
One sees that, because of the change in the trigger angle $\phi_s$ ($0\rightarrow\pi/2$), the cosine dependence of the background contribution has an opposite sign, as compared to the in-plane correlation. {\it We found the in-plane/out-of-plane effect!} 
To illustrate these results, plots are shown in Fig.~\ref{fig:2}, with  an appropriate choice of parameters. 

\section{Conclusions} 

In conclusion, the NeXSPheRIO code gives correct  qualitative behavior of the in-plane/out-of-plane effect. 
A simplified analytical one-tube model shows that this effect appears because, besides the contribution coming from the peripheral tube, additional contribution arises from the background elliptical flow, due to the multiplicity fluctuation. The latter is back-to-back  
($\Delta\phi=0\,,\,\pi$) in the case of in-plane triggers 
($\phi_S\sim0$) and rotated by $\pi/2$  
($\Delta\phi=-\pi/2\,,\,\pi/2$) in the case of 
out-of-plane triggers ($\phi_S\sim\pi/2$).

\section*{Acknowledgements}
We acknowledge funding from FAPESP and CNPq.

%\appendix
%\section{First Appendix} %Empty argument \section{} yields `Appendix'. 
%
%\section{Second Appendix}

\end{document}